\documentclass[journal,letterpaper]{IEEEtran_nssmic}
\usepackage{cite}      % Written by Donald Arseneau
\usepackage{graphicx}  % Written by David Carlisle and Sebastian Rahtz
\def\micron{{$\mu$m}}
\def\degC{{${}^\circ$C}}
\def\degree{{${}^\circ$}}

\begin{document}

\renewcommand{\thefootnote}{\fnsymbol{footnote}}

\onecolumn

\begin{flushright}
{\small
SLAC--PUB--10393\\
March, 2004\\}
\end{flushright}

\vspace{.8cm}

%%%%% Title and Author Information:
%%
\begin{center}
{\bf\large   
Performance of a Low Noise Front-end ASIC for Si/CdTe Detectors\\
in Compton Gamma-ray Telescope
\footnote{Work supported by
Department of Energy contract  DE--AC03--76SF00515, Grantin-Aid by Ministry of Education, Culture, Sports, Science and Technology of Japan (12554006, 13304014),
and ``Ground-based Research Announcement for Space Utilization'' promoted by Japan Space Forum.}}

\vspace{1cm}

Hiroyasu Tajima${}^1$, Tatsuya Nakamoto${}^2$, Takaaki Tanaka${}^{3,4}$, Shingo Uno${}^2$, 
Takefumi~Mitani${}^{3,4}$, 
Eduardo~do~Couto~e~Silva${}^1$, 
Yasushi~Fukazawa${}^2$, 
Tuneyoshi~Kamae${}^1$, 
Grzegorz~Madejski${}^1$, 
Daniel~Marlow,${}^5$
Kazuhiro~Nakazawa${}^3$, 
Masaharu~Nomachi,${}^6$
Yu~Okada${}^4$
and 
Tadayuki~Takahashi${}^{3,4}$

\medskip

${}^1$Stanford Linear Accelerator Center, Stanford University,
Stanford, CA  94309\\
${}^2$Department of Physics, Hiroshima University, 
Higashi-Hiroshima 739-8526, Japan\\
${}^3$Institute of Space and Astronautical Science, 
Sagamihara, Kanagawa 229-8510, Japan\\
${}^4$Department of Physics, University of Tokyo, 
Bunkyo-ku, Tokyo 113-0033, Japan\\
${}^5$Department of Physics, Priceton University, Princeton, NJ 08544\\
${}^6$Department of Physics, Osaka University, Toyonaka, Osaka 560-0043, Japan\\

\end{center}

\vfill

\begin{center}
{\bf\large   
Abstract }
\end{center}

\begin{quote}
Compton telescopes based on semiconductor technologies are being developed 
to explore the gamma-ray universe in an energy band 0.1--20~MeV, which is 
not well covered by the present or near-future gamma-ray telescopes. 
The key feature of such Compton telescopes is the high energy resolution 
that is crucial for high angular resolution and high background rejection 
capability.
The energy resolution around 1~keV is required to approach physical limit of the angular resolution due to Doppler broadening.
We have developed a low noise front-end ASIC (Application-Specific Integrated Circuit), VA32TA, to realize this goal for 
the readout of Double-sided Silicon Strip Detector (DSSD) and 
Cadmium Telluride (CdTe) pixel detector which are essential elements 
of the semiconductor Compton telescope.
We report on the design and test results of the VA32TA. 
We have reached an energy resolution of 1.3~keV (FWHM) for 60~keV and 122~keV 
at 0\degC\ with a DSSD and 1.7~keV (FWHM) with a CdTe detector.
\end{quote}

\bigskip

\noindent Index terms -- Analog integrated circuits, Gamma-ray detectors, Compton Camera, Silicon radiation detectors, Cadmium Telluride.

\vfill

%%%%%%%%%%%%%%%
%% Choose"Presented at," "Contributed to" for conference papers
%% or "Submitted to" for journal papers
%%%%%%%%%%%%%%%
\begin{center} 
{\it Contributed to} 
{\it IEEE Nuclear Science Symposium }\\
{\it Portland, Oregon USA}\\
{\it October 19--October 25, 2003} \\

%%OR\\

%%{\it Submitted to A Journal} (Spell out name of journal.)

\end{center}

\twocolumn

\clearpage

%%%%%%%%%%%%%%%%%%%%%%%%%%%%%%%%%%%%%%%%%%%%%%%%%%%%%%%%%%%%%
% Introduction
%%%%%%%%%%%%%%%%%%%%%%%%%%%%%%%%%%%%%%%%%%%%%%%%%%%%%%%%%%%%%
\section{Introduction}
% \PARstart{A}{demo} file is ....
% 
\PARstart{T}{he} gamma-ray universe in the energy band above 0.1~MeV provides 
a rich ground to study nucleosynthesis and physics of particle acceleration 
beyond thermal emission.
However, the energy band between 0.1~MeV and 100~MeV is poorly explored 
due to difficulties associated with the detection of such photons.
The Compton telescope COMPTEL~\cite{COMPTEL93} on board CGRO (Compton Gamma-Ray 
Observatory) demonstrated that a gamma-ray instrument based on the Compton 
scattering is useful for the detection of the gamma-ray in this energy band.
COMPTEL provided us rich information on a variety of gamma-ray emitting 
objects either in continuum and line emission. 
The continuum sources include spin-down pulsars, stellar black-hole candidates,
supernovae remnants, interstellar clouds, active galactic nuclei (AGN), 
gamma-ray bursts (GRB) and solar flares. 
Detection has also been made of the nuclear gamma-ray lines from 
${}^{26}$Al (1.809~MeV), ${}^{44}$Ti (1.157~MeV), and 
${}^{56}$Co (0.847 and 1.238~MeV). 

Although COMPTEL performed very well as the first Compton telescope 
in space for MeV gamma-ray astrophysics, it suffered severely 
from large background, poor angular resolution, and complicated image 
decoding~\cite{Knodlseder96}.
In 1987, T. Kamae {\it et al}. proposed a new Compton telescope based 
on a stack of silicon strip detectors (SSD)~\cite{Kamae87,Kamae88}.
This technology presents very attractive possibilities to overcome 
the weaknesses of COMPTEL as described later in this document.
This idea of using silicon strip detectors stimulated new proposals 
for the next generation Compton telescope~\cite{Takahashi,Takahashi03-SGD,
MEGA,Milne}.
%Scientific objectives of such telescopes include:
%\begin{itemize}
%\item Studies of nucleosynthesis in supernovae, novae, massive stars 
%and galactic supernovae via detection of nuclear gamma-ray lines.
%\item AGN: High-energy emission of blazars (AGNs) peaks in 1--20~MeV range. 
%Detection of $e^+/e^-$ annihilation line would provide information about 
%the innermost region of the central engine.
%\item GRB: The spectrum peaks at 0.1--0.3~MeV range. The gamma-ray emission 
%mechanism can be studied by measuring A detailed description of the SSD 
%system and the noise and energy resolution measurements can be found 
%elsewhere~\cite{Tajima02}.
%photon polarization.
%\end{itemize}

Recently, a new semiconductor detector based on Cadmium Telluride (CdTe) 
emerged as a promising detector technology for detection of MeV 
gamma-rays~\cite{Takahashi01,Nakazawa03}.
Taking advantage of significant development in CdTe technology, we are 
developing a new generation of Compton telescopes, the SGD (Soft Gamma-ray 
Detector)~\cite{Takahashi03-SGD} onboard the NeXT (New X-ray Telescope) 
mission proposed at ISAS (Institute of Space and Astronautical Science) as a successor of the Astro-E2, and the SMCT (Semiconductor Multiple-Compton Telescope).
The NeXT/SGD is a hybrid semiconductor gamma-ray detector which consists of 
silicon and CdTe detectors to measure photons in a wide energy 
band (0.05--1 MeV);
the silicon layers are required to improve the performance at a lower energy 
band ($<$0.3 MeV).
The NeXT/SGD is a Compton telescope with narrow field of view (FOV), which utilizes
Compton kinematics to enhance its background rejection capabilities.
The SMCT will have a wider energy band (0.1--20 MeV) and a wide field 
of view ($\sim$60\degree).
Excellent energy resolution is the key feature of the NeXT/SGD and SMCT,
allowing to achieve both high angular resolution 
and good background rejection capability.
It is worthwhile mentioning an additional capability of the NeXT/SGD and SMCT,
their ability to measure $\gamma$-ray polarization, which opens up a new
window to study properties of astronomical objects.

A low noise front-end ASIC is a fundamental element of semiconductor Compton 
telescopes.
In this paper, we report on the design of the front-end ASIC and test results 
of prototype systems.

%MEGA [6] project is well advanced in R&D but less aggressive in terms of the 
%angular resolution (2.6o FWHM @ 2~MeV), sensitivity and energy range. 
%The NRLfs proposal [7] is more aggressive but early in the development 
%phase.

%%%%%%%%%%%%%%%%%%%%%%%%%%%%%%%%%%%%%%%%%%%%%%%%%%%%%%%%%%%%%
% Multi-Compton Technique.
%%%%%%%%%%%%%%%%%%%%%%%%%%%%%%%%%%%%%%%%%%%%%%%%%%%%%%%%%%%%%
\section{Multiple-Compton technique}
\begin{figure}[bth]
   \begin{center}
   \includegraphics[height=6cm]{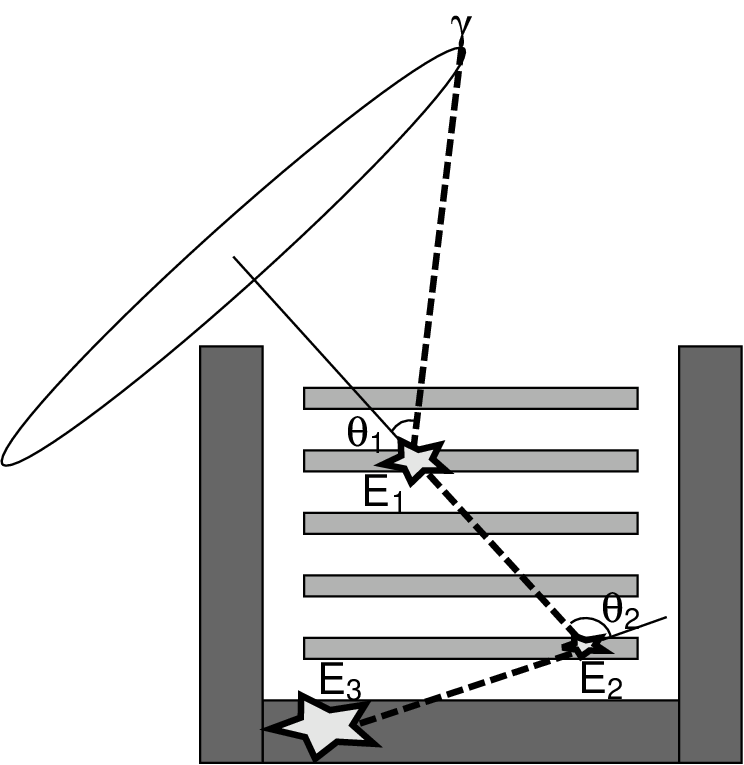}
   \end{center}
\vspace*{-0.4cm}
   \caption[Concept of Multiple-Compton technique.] 
%>>>> use \label inside caption to get Fig. number with \ref{}
   { \label{fig:schematic} Concept of the Multiple-Compton technique.}
   \end{figure}

A stack of many thin scatterers is used in the multiple-Compton
technique~\cite{Kamae87}, 
which can accommodate more elements in the stack, 
thereby increasing the  detection efficiency 
%allows increased detection efficiency by stacking more elements, 
while maintaining the ability to record individual Compton scatterings.  
Fig.~\ref{fig:schematic} illustrates the case for two Compton scatterings 
and one photoelectric absorption.
In this situation, the scattering angles $\theta_1$ and $\theta_2$ can both 
be obtained from the recoil electron energies from the relations
%\begin{displaymath}
\begin{eqnarray}
\cos\theta_1 &=& 1+\frac{m_ec^2}{E_1+E_2+E_3}-\frac{m_ec^2}{E_2+E_3},\\ 
\cos\theta_2 &=& 1+\frac{m_ec^2}{E_2+E_3}-\frac{m_ec^2}{E_3},
\label{eq:kinematics}
\end{eqnarray}
%\end{displaymath}
where $E_1$, $E_2$ and $E_3$ are the energy deposited in each photon 
interaction.  
Note that $\theta_2$ can be also reconstructed from the hit positions of 
the three interactions.
The order of the interaction sequences, hence the correct energy and 
direction of the incident photon, can be reconstructed by examination 
of this constraint for all possible sequences.  
This over-constraint also provides stringent suppression of random 
coincidence backgrounds.
The direction of the incident photon can be confined to be on the surface 
of a cone determined from $\theta_1$ and the first two interaction positions.
Precise energy resolution\footnote{Here, energy resolution includes all sources such as noise, gain uncertainty and fluctuation of electron-hole pair generation in the detector. The noise is the dominant source in our application.} for the recoil electron is the critical feature 
in the design of Compton telescope since the angular resolution 
of the incident photon, and the background rejection capability are 
determined by the Compton kinematics.
We have demonstrated that the Compton technique can be used 
to measure the photon direction and polarization using a stack of SSD 
and CdTe detectors as described in~\cite{Tajima03, Mitani03}.

%-------------
%-------------

\begin{figure}[tbh]
   \begin{center}
   \includegraphics[height=5cm]{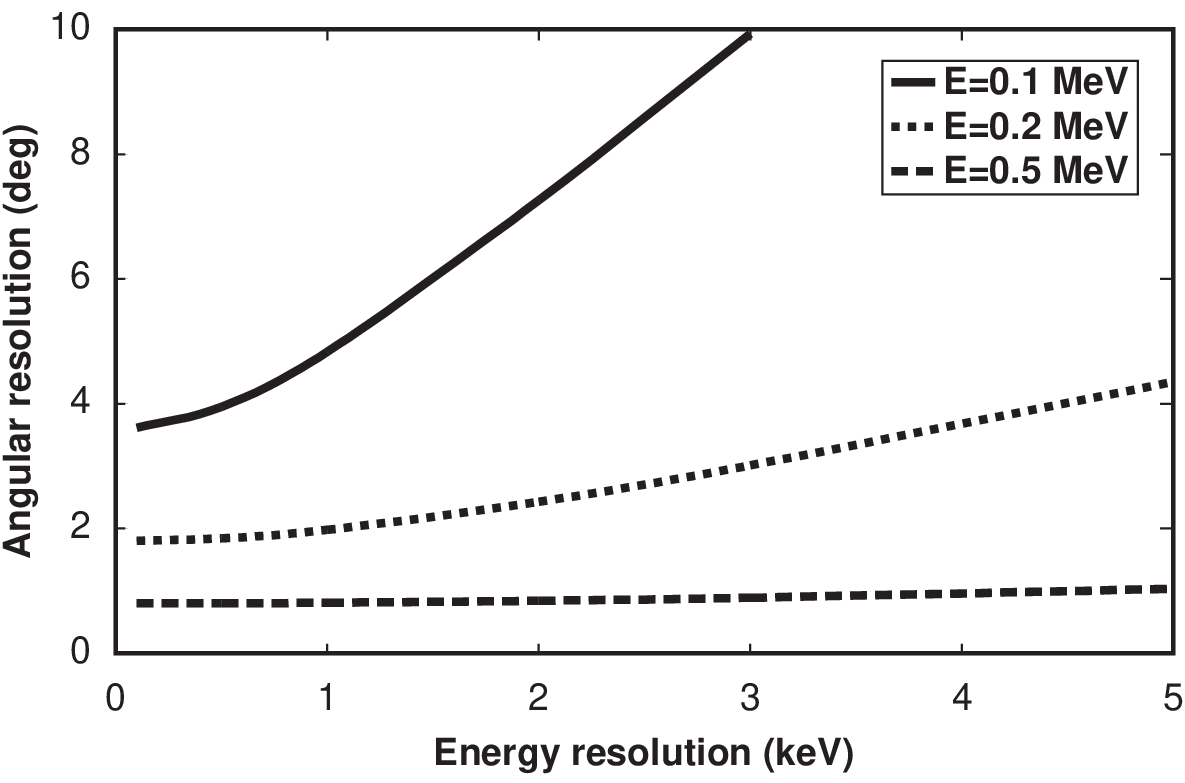}
   \end{center}
\vspace*{-0.4cm}
   \caption[Angular resolution as a function of 
the energy resolution for photons scattering in silicon with incident 
energies of 100, 200 and 500 keV at $\cos\theta=0.5$.] 
%>>>> use \label inside caption to get Fig. number with \ref{}
   { \label{fig:angle-energy} Angular resolution as a function of 
the energy resolution for photons scattering in silicon with incident 
energies of 100, 200 and 500 keV at $\cos\theta=0.5$.}
   \end{figure}

Fig.~\ref{fig:angle-energy} shows the angular resolution as a function of 
the energy resolution for photons scattering in silicon with incident 
energies of 100, 200 and 500 keV at $\cos\theta=0.5$ where $\theta$ is the polar angle of the Compton scattering.
Due to Doppler braodening effect, the energy resolution below 1~keV does not 
necessarily result in better angular resolution for the incident photon energy above 100~keV.
In order to achieve the energy resolution of 1.0~keV (FWHM, Full-Width at Half 
Maximum) for the silicon, a low noise ASIC with an equivalent noise charge 
(ENC) of better than 120~$e^-$ (rms) is required.

%The DSSD is a primary candidate for the scatterer because Compton scattering 
%dominates over other processes in sub-MeV range and the Doppler-broadening 
%effect is small in silicon.
%Such double-sided devices are required to provide two dimensional measurement 
%of the interaction position.
%Weak stopping power of the silicon detector at the upper end of the energy 
%band can be supplemented by employing the CdTe detectors in the back portion 
%of the layers.
%The DSSD layers measure the recoil electron energy of the first Compton 
%scattering with maximal energy resolution while the CdTe layers enhance 
%the probability of Compton scattering and/or photoelectric absorption 
%in the back section. 
%At higher energies, silicon becomes more or less transparent and CdTe 
%detector becomes the dominant detector.
%Pair creation can also be utilized to measure both the energy and 
%direction of gamma-rays above several MeV.
%However, the efficiency is significantly reduced above 20 MeV since 
%the gamma-ray cannot be completely absorbed.

%-------------
%-------------

%-------------
   \begin{figure}[tbh]
   \begin{center}
   \begin{tabular}{c}
   \includegraphics[height=2.2cm]{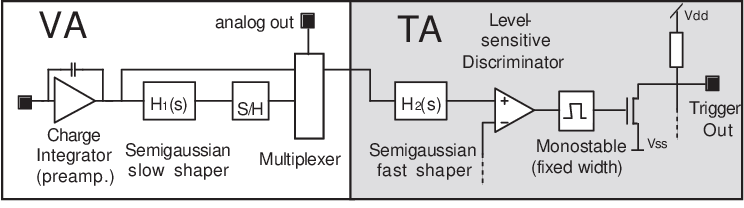}
   \end{tabular}
   \end{center}
\vspace*{-0.4cm}
   \caption[Block diagram of VA32TA preamplifier VLSI.] 
%>>>> use \label inside caption to get Fig. number with \ref{}
   { \label{fig:VA32TA} Block diagram of VA32TA front-end VLSI.}
   \end{figure} %
%-------------

%Possible instrument configurations and expected performance are discussed 
%in ref~\citenum{Takahashi02}.

%%%%%%%%%%%%%%%%%%%%%%%%%%%%%%%%%%%%%%%%%%%%%%%%%%%%%%%%%%%%%
% low noise ASIC
%%%%%%%%%%%%%%%%%%%%%%%%%%%%%%%%%%%%%%%%%%%%%%%%%%%%%%%%%%%%%
\section{Low Noise ASIC}
%A detailed description of the SSD system and the noise and energy resolution 
%measurements can be found elsewhere~\cite{Tajima02}.

\begin{figure}[tbh]
   \begin{center}
   \includegraphics[height=7.5cm]{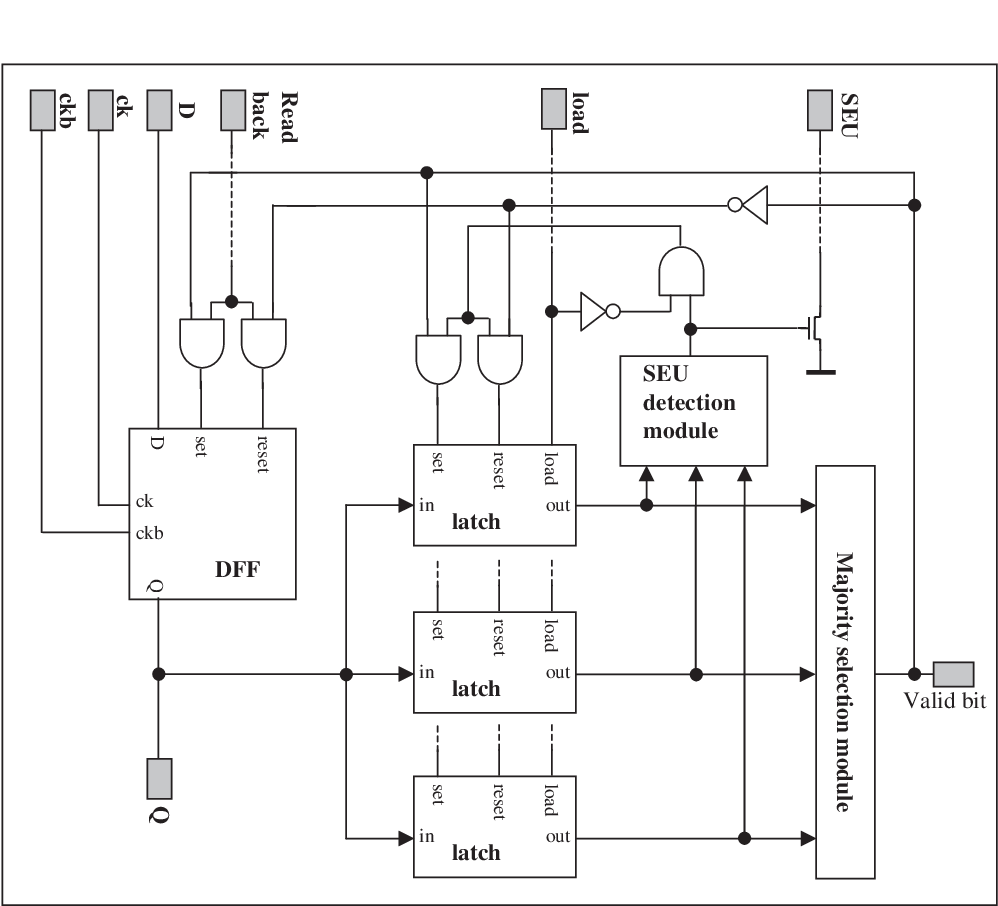}
   \end{center}
\vspace*{-0.4cm}
   \caption[A circuit diagram of the majority selector logic.] 
%>>>> use \label inside caption to get Fig. number with \ref{}
   { \label{fig:majority} A circuit diagram of the majority selector logic.}
   \end{figure}

We have developed the VA32TA front-end ASIC based on the design of the VA32C 
amplifier VLSI (Very-Large-Scale Integration) and the TA32C trigger VLSI that are originally developed 
by Ideas.\footnote{Ideas ASA, Veritaveien 9, 1363 Hovik, Norway}
A detailed description of Viking-architecture (VA) chip is given 
elsewhere~\cite{VA91,VA94}.
The VA32TA is fabricated in the AMS\footnote{Austria Micro Systems AG, A-8141, Schloss Premst\"{a}tten, Austria} 0.35 $\mu$m technology with epitaxial 
layer, which is measured to be radiation tolerant up to 20~MRad or more 
\cite{Yokoyama01}.
The epitaxial layer with careful grounding improved the tolerance against 
Single-Event Latch-up to values greater than 170~MeV/\micron${}^2$~\cite{Korpar03}.
A VA32TA consists of 32 channels of signal-readout.  
Each channel includes a charge sensitive preamplifier, slow CR-RC 
shaper, sample/hold and analog multiplexer chain (VA section), and fast 
shaper and discriminator chain (TA section) as illustrated in the block diagram 
of Fig.~\ref{fig:VA32TA}.
The front-end MOSFET geometry for the preamplifier was originally optimized for small capacitance load in the AMS 1.2 \micron\ process.
We did not re-optimize the geometry in the 0.35~\micron\ process to minimize the development time and risk since the main objective of this development is achieving low noise while integrating the VA and TA designs.
The FET geometry will be optimized in the 0.35~\micron\ process in the next development cycle where the low power consumption is the main issue.
Expected noise performance is $(45+19\times C_d)/\sqrt{\tau}\;e^-$ (rms) in 
ENC, where $C_d$ is the load capacitance in pF and $\tau$ is the peaking 
time in $\mu$s, which can be varied from 1 to 4 $\mu$s.
The $1/f$ noise is negligible.
Feedback resistors for the preamplifier, as well as slow and fast shapers 
are realized with MOSFETs.
Gate voltages of the feedback MOSFETs are controlled by internal 
DACs (Digital-to-Analog Converters) on chip.
Bias currents for various components are also controlled by the internal DACs.
Threshold levels can be adjusted for each channel using individual DACs 
to minimize threshold dispersion.
A 200-bit register is required to hold the values for all internal DACs.
Majority selector logic circuitry has been utilized for these registers 
to ensure the tolerance against Single-Event Upset (SEU), which is 
important for space applications.
This majority selector circuitry uses three flip-flops for each bit and takes 
a majority of the three when they become inconsistent as shown in 
Fig.~\ref{fig:majority}.
This logic also generates a signal when such inconsistencies are detected.
The SEU tolerance of single latch is measured to be greater than 
70~MeV/\micron${}^2$~\cite{Korpar03}.
Two latches need to be upset at the same time to permanently upset a register bit.
%-------------
%-------------

%%%%%%%%%%%%%%%%%%%%%%%%%%%%%%%%%%%%%%%%%%%%%%%%%%%%%%%%%%%%%
% noise performance
%%%%%%%%%%%%%%%%%%%%%%%%%%%%%%%%%%%%%%%%%%%%%%%%%%%%%%%%%%%%%
\section{ Noise Performance }
We have developed prototype modules for a low noise Double-sided Silicon 
Strip Detector (DSSD) system in order to evaluate noise sources.
A low noise DSSD system consists of a DSSD, an RC chip and a VA32TA front-end VLSI chip. 
To keep the strip yield close to 100\% and eliminate polysilicon bias 
resistor (a possible noise source), the DSSD does not employ an integrated 
AC capacitor.
We have produced 300~\micron\ thick DSSDs with a strip length of 2.56~cm, 
a strip gap of 100~\micron\ and a strip pitch of 400~\micron.
The C-V measurement indicates a depletion voltage of 65~V, 
therefore the following measurements are performed with a 70~V bias voltage.  
%No junction breakdown was observed up to 200~V bias voltage.  
The leakage current is 0.5~nA/strip at 20\degC\ and 0.05~nA/strip at 0\degC.
The strip capacitance is measured to be $6.3\pm0.2$~pF.
The RC chip provides detector bias voltage via polysilicon bias resistors, 
as well as AC-coupling between strips and preamplifier channel inputs.  
A resistance value of 1 G$\Omega$ is chosen for the bias resistor in order 
to minimize thermal noise without compromising production stability.
We have assembled three prototype modules: one consists of a single-sided SSD, an RC chip and a VA32TA (AC module);  another consists of a single-sided SSD and a VA32TA (DC module); the other consists of a DSSD, RC chips and VA32TAs (DSSD module). The RC chips are used on only ohmic side of the DSSD module to minimize the noise on the junction side.
The two single-sided SSD modules are used to study the effect of the RC chip on the noise performance.
The performance of the junction side of the DSSD module is measured to be identical to the DC module.
A detailed description of the SSD system and the noise and energy resolution 
measurements can be found elsewhere~\cite{Tajima02}.

We have taken into account the following noise sources in our analysis to 
optimize the component parameters:
\begin{itemize}
\item Preamplifier noise is characterized as $(0.37+0.16\times C_d)/\sqrt{\tau}$~keV (FWHM) for silicon detector. 
Capacitance load for the DC module is estimated to be 7~pF including the strip capacitance and the parasitic capacitance due to wire bonds.
Capacitance load from the RC chip is estimated to be 3~pF from the geometry.
Noise level varies from 0.8~keV (DC, $\tau=4$ $\mu$s) to 1.4~keV (AC, 
$\tau=2$ $\mu$s).
\item Shot noise due to the leak current ($I_d$~nA) is characterized as $0.87\sqrt{I_d\times \tau}$~keV (FWHM).
Measured leak current is used to calculate the noise.
Noise level varies from 0.3~keV (0\degC, $\tau=2$ $\mu$s) to 1.2~keV 
(20\degC, $\tau=4$ $\mu$s).
\item Thermal noise due to the bias resistor ($R_B$~G$\Omega$) in the RC chip is characterized as $0.20\sqrt{\tau/R_B}$~keV (FWHM).
Noise level varies from 0.3~keV ($\tau=2$~$\mu$s) to 0.4~keV ($\tau=4$~$\mu$s).
\end{itemize}
We have also considered the thermal noise from the implant resistance of 
the RC chip and found it is less than 0.1~keV, hence considered negligible.
The noise analysis indicates that the total noise is fairly independent of the peaking time beyond 2~$\mu$s because the preamplifier noise and other noise sources show opposite peaking time dependence and compensate each other.
\begin{table*}[bth]
\caption{Noise and energy resolution measured for  the prototype modules 
in the AC and DC configurations at temperatures of 0\degC\ and 20\degC\ 
and peaking times of 2~$\mu$s and 4~$\mu$s.}
\label{tab:resolution}
\begin{center}       
\begin{tabular}{|c|r|r||r|r||r|r|} 
%% use of \rule[]{}{} below opens up each row
\hline
\rule[-1ex]{0pt}{3.5ex}   &  & Peaking & \multicolumn{2}{c||}{ Noise (FWHM)} &  \multicolumn{2}{c|}{Energy resolution (FWHM)} \\
\cline{4-7}
\rule[-1ex]{0pt}{3.5ex} \raisebox{1.5ex}[0pt]{Configuration} & \raisebox{1.5ex}[0pt]{Temperature}  &  \multicolumn{1}{c||}{time} & \multicolumn{1}{c|}{Expected} &  \multicolumn{1}{c||}{Measured} & \multicolumn{1}{c|}{60~keV (${}^{241}$Am)} & \multicolumn{1}{c|}{122~keV (${}^{57}$Co)}\\
\hline\hline
\rule[-1ex]{0pt}{3.5ex} &  & 2 $\mu$s  & 1.6~keV & 1.7~keV & 1.9~keV & 2.1~keV\\
\rule[-1ex]{0pt}{3.5ex}  & \raisebox{1.5ex}[0pt]{20\degC} & 4 $\mu$s  & 1.6~keV & 1.7~keV & 2.1~keV & 2.3~keV\\
\cline{2-7}
\rule[-1ex]{0pt}{3.5ex}  \raisebox{1.5ex}[0pt]{AC} &  & 2 $\mu$s  & 1.4~keV & 1.6~keV & \hspace*{0.8cm}1.8~keV  & \hspace*{0.8cm}1.9~keV\\
\rule[-1ex]{0pt}{3.5ex} & \raisebox{1.5ex}[0pt]{0\degC} & 4 $\mu$s  & 1.1~keV & 1.4~keV & 1.7~keV & 1.8~keV\\
\hline
\rule[-1ex]{0pt}{3.5ex} &  & 2 $\mu$s  & 1.1~keV & 1.2~keV & 1.6~keV & 1.6~keV\\
\rule[-1ex]{0pt}{3.5ex} \raisebox{1.5ex}[0pt]{DC} & \raisebox{1.5ex}[0pt]{0\degC} & 4 $\mu$s  & 0.9~keV & 1.0~keV & 1.3~keV & 1.3~keV\\
\hline
\end{tabular}
\end{center}
\end{table*}
%-------------
   \begin{figure*}[tbh]
   \begin{center}
   \begin{tabular}{ll}
   \includegraphics[height=4.6cm]{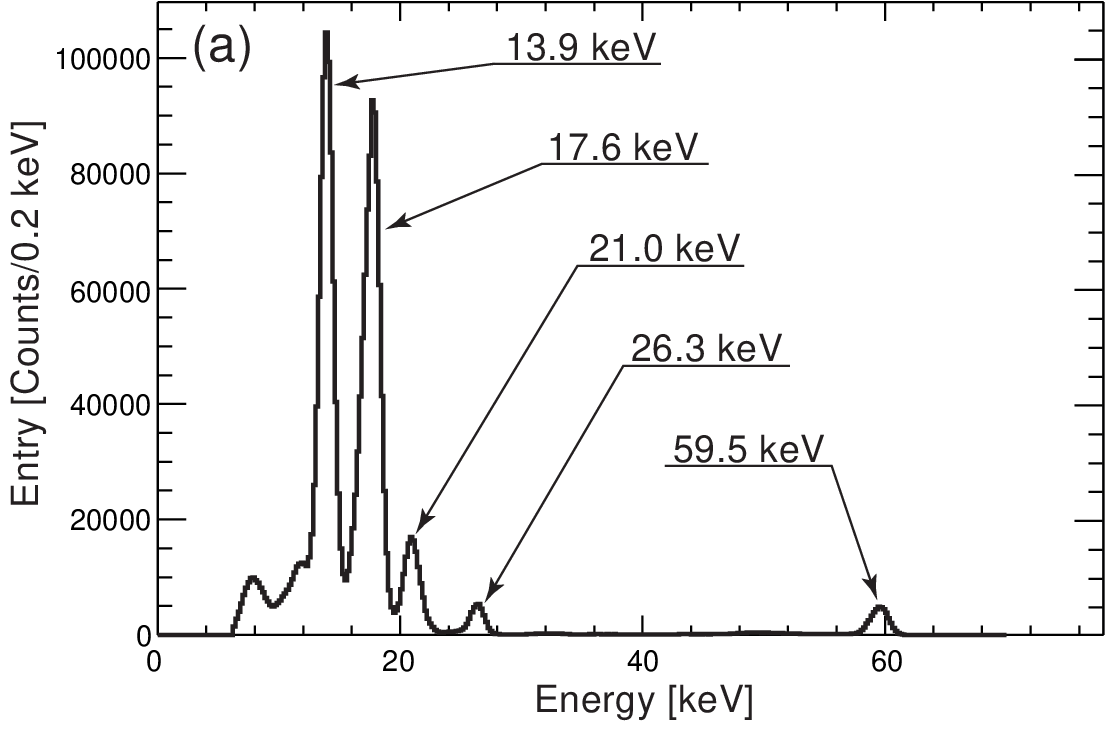} \hspace*{0.8cm} &
   \includegraphics[height=4.6cm]{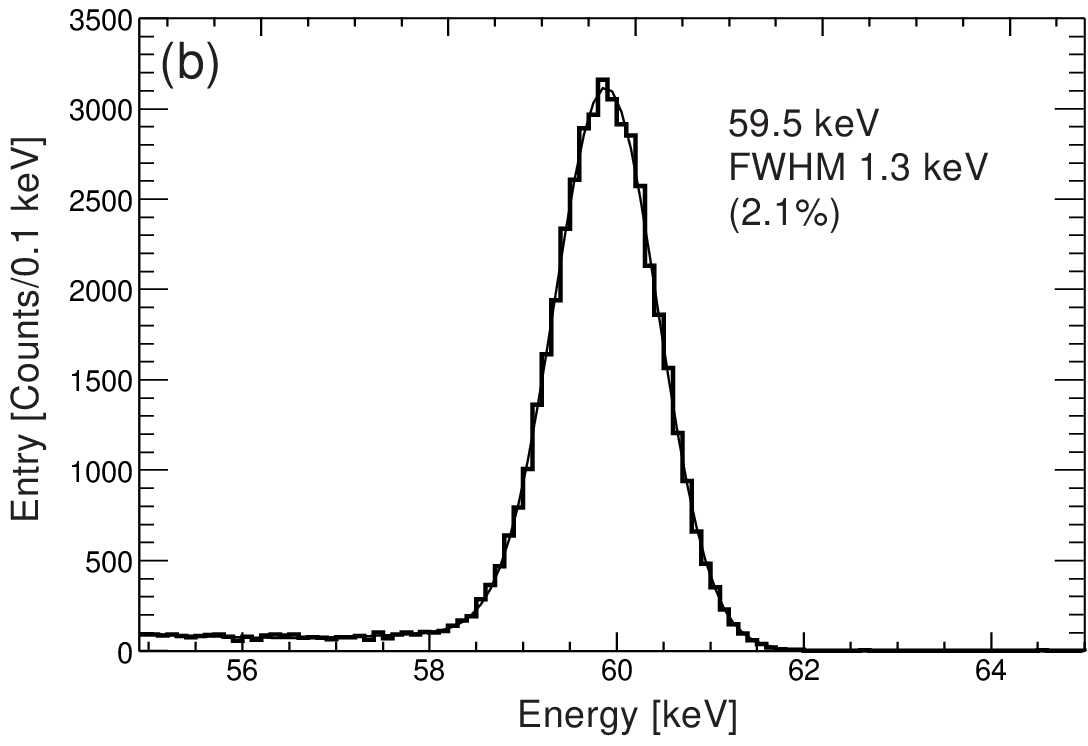} % or 5.1cm
   \end{tabular}
   \end{center}
\vspace*{-0.4cm}
   \caption[(a) Energy spectrum of ${}^{241}$Am measured by the DC module 
at 0\degC. 
(b) Magnified view around the 59.5~keV $\gamma$-ray peak. The curve 
represents the fit result described in the text.] 
%>>>> use \label inside caption to get Fig. number with \ref{}
   { \label{fig:Am} (a) Energy spectrum of ${}^{241}$Am measured by the DC 
module at 0\degC. (b) Magnified view around the 59.5~keV $\gamma$-ray peak. 
The curve represents the fit result described in the text.}
   \end{figure*} 
%-------------
%-------------
   \begin{figure*}[tbh]
   \begin{center}
   \begin{tabular}{ll}
   \includegraphics[height=4.6cm]{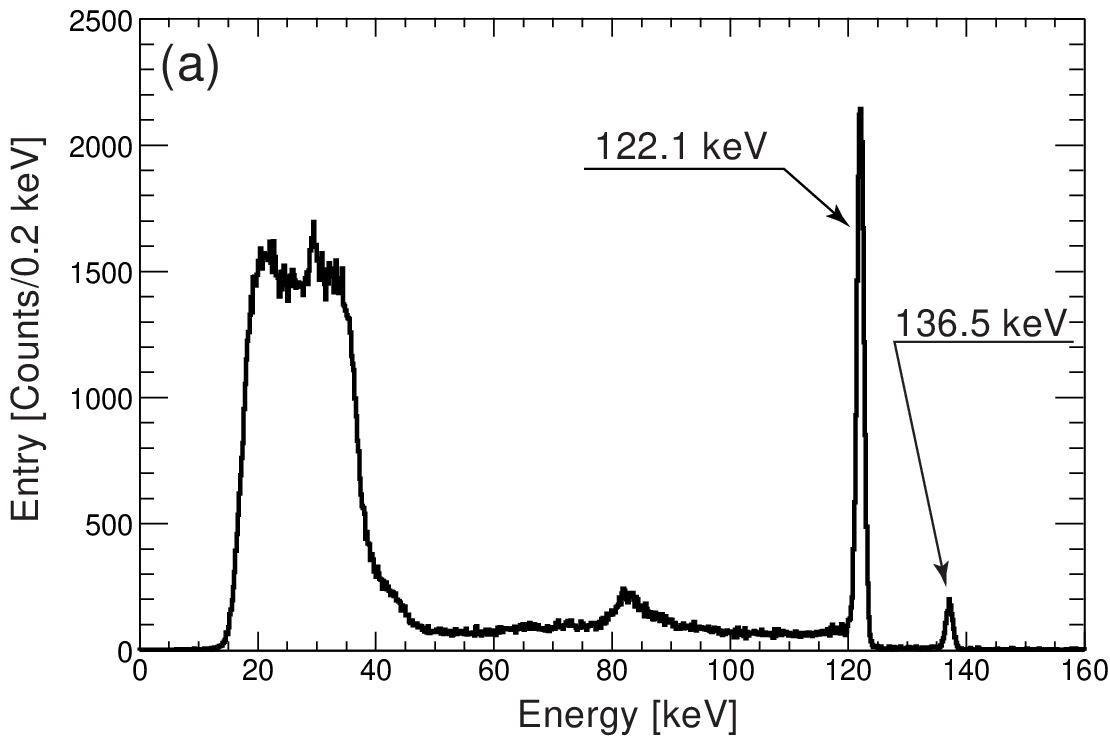} \hspace*{0.8cm} &
   \includegraphics[height=4.6cm]{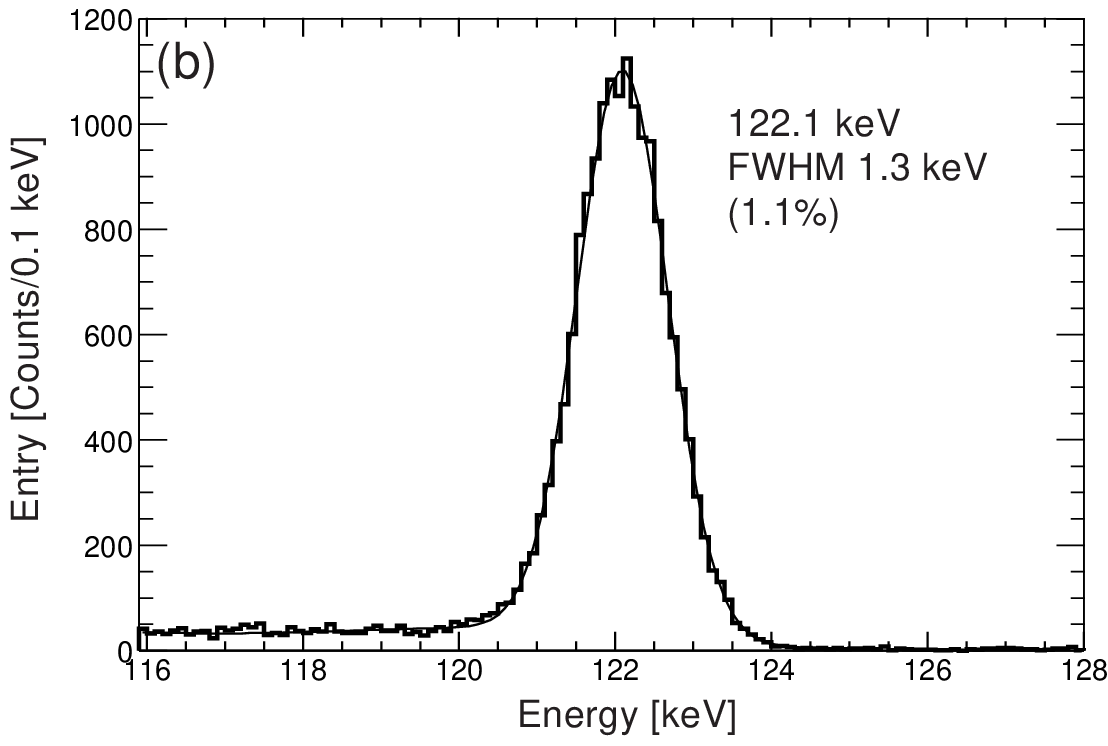} % or 5.1cm
   \end{tabular}
   \end{center}
\vspace*{-0.4cm}
   \caption[(a) Energy spectrum of ${}^{57}$Co measured by the DC module 
at 0\degC. (b) Magnified view around the 122.1~keV X-ray peak. 
The curve represents the fit result described in the text.] 
%>>>> use \label inside caption to get Fig. number with \ref{}
   { \label{fig:Co} (a) Energy spectrum of ${}^{57}$Co measured by the DC 
module at 0\degC. (b) Magnified view around the 122.1~keV $\gamma$-ray peak. 
The curve represents the fit result described in the text.}
   \end{figure*} 
%-------------

The noise performance of the prototype system is measured at temperatures of 
0\degC\ and 20\degC\ and at peaking times of 2~$\mu$s and 4~$\mu$s.  
Varying these parameters is useful to differentiate the noise contributions.  
The absolute gain of the system is calibrated using the $\gamma$-ray spectra 
described below and is approximately 80--100 mV/fC.
We obtain noise performance of 1.0~keV at 0\degC\
and $\tau = 4$~$\mu$s, which is in a good agreement with the expected value
derived from the known noise performance of the VA32TA and the measured strip
capacitance.
Table~\ref{tab:resolution} summarizes the measurement results and compares 
them with calculation results.
This table does not include the result for the DC configuration at 20\degC\ 
since it cannot be operated at optimum condition due to the effect of leak 
current from the SSD flowing into the preamplifier.

These results confirm that the shot noise due to the leakage current becomes 
negligible at 0\degC.
In the DC configuration, the measured and expected noise values are in a good 
agreement for the measurements at either peaking time, which demonstrates 
that the noise sources from the SSD and the VA32TA are well understood.
On the other hand, slight disagreement between measurement and calculation 
is observed for the AC configuration at 0\degC, indicating additional noise 
sources in the RC chip which are not accounted for in the noise analysis.
At 20\degC\ the shot noise from the leak current becomes large and 
the disagreement becomes less significant.
Further studies are required to identify the origin of the excess noise 
observed in the RC chip in this measurement.

%%%%%%%%%%%%%%%%%%%%%%%%%%%%%%%%%%%%%%%%%%%%%%%%%%%%%%%%%%%%%
% Gamma-ray energy resolution
%%%%%%%%%%%%%%%%%%%%%%%%%%%%%%%%%%%%%%%%%%%%%%%%%%%%%%%%%%%%%
\section{ $\gamma$-ray energy resolution}
The energy resolution for $\gamma$-rays is investigated
using the 59.54~keV $\gamma$-ray line from ${}^{241}$Am and the
122.06~keV $\gamma$-ray line from ${}^{57}$Co.  
The absolute gain is calibrated for each channel using the same $\gamma$-ray 
lines in such way that all channels give the nominal peak height.
We also correct for the gain dependence on the ``common mode shift" defined 
as the baseline shift common to all channels in one chip.\footnote{The 
common mode shift is calculated and subtracted for each event taking 
the average of the pulse height for every channel, excluding the channels 
with signal.}
We attribute the origin of this dependence to the nonlinearity of the 
amplifier gain.
A correction of approximately 0.2~keV is made for a 1~keV common mode shift.
This correction is critical to operate the system at 20\degC, since 
the magnitude of the common mode shift is 6--10~keV at 20\degC\ 
and 1.6~keV at 0\degC.
Larger common mode shift is observed for the longer peaking time at 20\degC, 
implying on a low frequency nature of the common mode shift.
It should be noted that the peaking time of the fast shaper in the TA section 
is 0.3~$\mu$s, which helps to suppress the contribution from common mode shift.

Figs.~\ref{fig:Am} (a) and \ref{fig:Co} (a) show the sum of the  energy 
spectra for all channels, except for the first and last strips where we 
observe larger noise.
We observe a clear Compton edge just below 40~keV in the ${}^{57}$Co spectrum.
In order to obtain the energy resolution, we fit the energy spectrum to 
a sum of two Gaussian functions to describe the peak and one threshold type 
function\footnote{This function is expressed as $f(E) = p_1\cdot (E_0-E)^{p_2}
\cdot e^{-p_3(E_0-E)},\;\;(E<E_0)$.} to describe the continuum spectrum.
The second Gaussian is introduced to take into account the effect of the missing charge due to escaped electrons and charge sharing between strips.
The mean value of the second Gaussian is shifted by 1--2~keV due 
to missing charge.
The fraction of events in the second Gaussian is in general less than 15\%.
Figs.~\ref{fig:Am} (b) and \ref{fig:Co} (b) show the magnified energy 
spectra around the peak and the fit results.

Table~\ref{tab:resolution} summarizes the results of the energy resolution 
measurements.
Energy resolution is worse than the intrinsic noise performance of the system.
The fluctuation of the electron-hole pair generation degrades the energy resolution to 1.1~keV from 1.0~keV at 122~keV.
The remaining contribution is predominantly due to the gain stability or uncertainty.
%Gain uncertainty can also explain the fact that the energy resolution is 
%worse for higher energy.

This result demonstrates that the energy resolution around 1~keV is 
an achievable goal, considering the planned improvements discussed below.

\begin{figure}[tbh]
   \begin{center}
   \includegraphics[height=4.6cm]{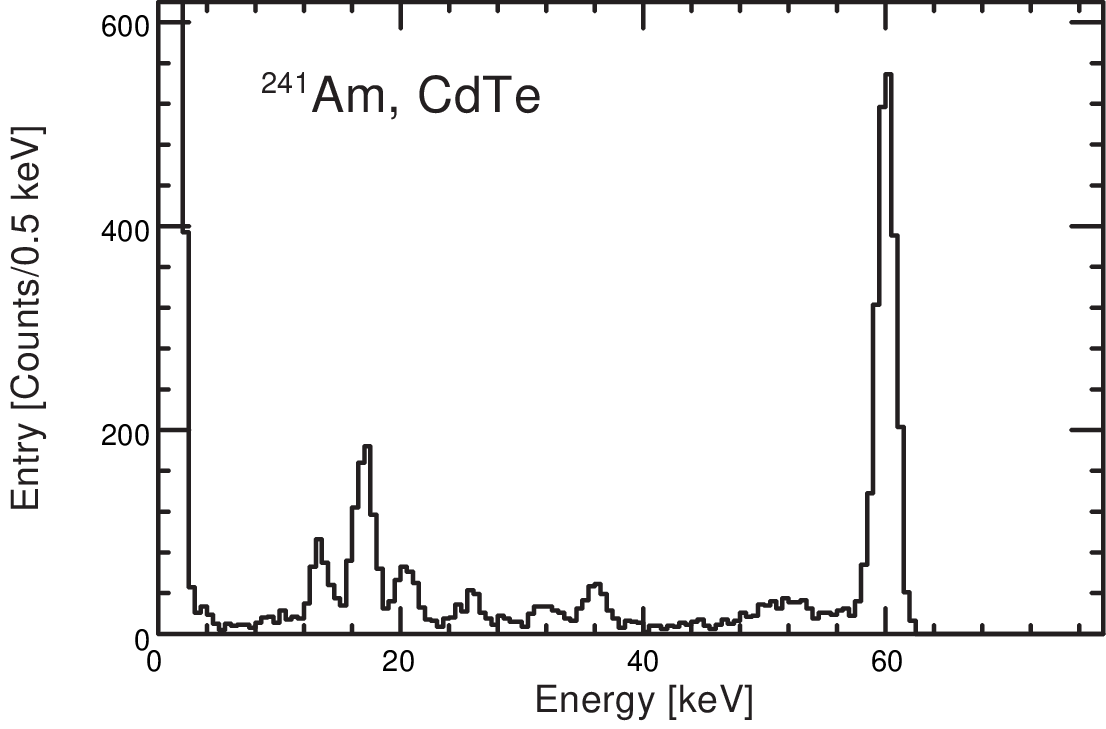} % or 5.1cm
   \end{center}
\vspace*{-0.4cm}
   \caption
   { \label{fig:CdTe-spectrum} The ${}^{241}$Am energy spectrum for 
a CdTe pixel detector. }
   \end{figure}

The VA32TA performance for a CdTe detector is also evaluated using an $8\times
8$ array of 2 mm${}\times$ 2 mm CdTe pixels.
The thickness of the detector is 0.5 mm.
Each pixel is connected to VA32TA via a fan-out board.
Capacitance and leakage current of each pixel are 1~pF and a few pA at 0\degC.
Low leakage current is realized by employing a guard ring to absorb leakage 
current from the detector edge~\cite{Nakazawa03}.
Noise is measured to be 1.5 keV (FWHM).
Fig.~\ref{fig:CdTe-spectrum} shows the ${}^{241}$Am energy spectrum 
at a bias voltage of 600~V.
We obtain an energy resolution of 1.7 keV (FWHM).
%-------------
%-------------

\section{Conclusions and future prospect}
We have developed a low noise front-end ASIC, VA32TA, 
for semiconductor Compton telescopes.
Intrinsic noise performance with a silicon strip detector is measured 
to be 1.0~keV (FWHM) at 0\degC\ in the DC configuration, which is 
in good agreement with the analytically calculated noise value of 0.9~keV.
The energy resolution is measured to be 1.3~keV (FWHM) for silicon strip 
detector and 1.7~keV for a CdTe pixel detector, demonstrating that 
the energy resolution around 1~keV is within our reach.

We plan to develop a full-size DSSD module based on the information obtained 
with the present prototype.
The size of the DSSDs will be 5 cm $\times$ 5 cm, which is the largest size 
possible with 4-inch wafers.
This size translates into twice the capacitance load for amplifier in the 
DC configuration.
Since the noise performance is dominated by the amplifier noise, improvements 
in the amplifier performance and reduction of the load capacitance is 
essential to achieve 1~keV energy-resolution.
A low power version of the VA32TA, VA32TALP, is being developed by further 
optimizing the front-end MOSFET geometry in the 0.35~$\mu$m 
process.
%\footnote{The current MOSFET geometry is optimized in the AMS 1.2~$\mu$m process.}
The smaller feature size of the 0.35~\micron\ process enable us to reduce the FET channel length ($L$), which in general improves the transconductance of the FET, i.e. noise performance \cite{OConnor02}.
Simulation studies indicate that ENC of $38+12\times C_d$ $e^-$ (rms) 
can be achieved with $L=0.4$ \micron\ at a peaking time of 4 $\mu$s despite 
the fact that the power consumption is reduced significantly to 0.2~mW/channel 
from 2~mW/channel.
Approximately 10--20\% improvement on the noise performance is expected 
by moving to $L=0.4$ \micron\ from $L=1.2$ \micron.
We also plan to employ a thicker DSSD to reduce the capacitance to the 
backside.

In conclusion, we have demonstrated that energy resolution of silicon and 
CdTe detectors are as low as 1.3~keV and 1.7~keV, respectively, with 
the possibility of the future improvements.
Such an ultra low noise ASIC presents great possibilities for future 
gamma-ray telescopes, such as the NeXT/SGD and SMCT.

%\section*{Acknowledgment}
%This work has been carried out under support of U.S. Department of Energy, 
%contract DE-AC03-76SF00515,
%Grantin-Aid by Ministry of Education, Culture, Sports, Science and Technology 
%of Japan (12554006, 13304014),
%and ``Ground-based Research Announcement for Space Utilization'' promoted 
%by Japan Space Forum.

%\IEEEtriggeratref{10}
\bibliographystyle{IEEEtran.bst}
%\bibliographystyle{elsart-num}   %>>>> makes bibtex use mybib.bst
%\bibliographystyle{spiebib}   %>>>> makes bibtex use spiebib.bst
%\bibliography{mybib} 
\bibliography{mybib}

\end{document}